\documentclass[prl,aps,twocolumn,showpacs]{revtex4}
\begin{document}

\def\bra#1{\left\langle#1\right|}
\def\ket#1{\left|#1\right\rangle}


\title{Enscription of Quantum Texts}
\author{Randall Espinoza, Tom Imbo and Paul Lopata}
\affiliation{Department of Physics, 845 W. Taylor St., University of Illinois at Chicago, Chicago, IL 60607-7059}

\begin{abstract}
We investigate an entangled deformation of the deterministic quantum cloning process, called enscription, that can be applied to (certain) sets of distinct quantum states which are not necessarily orthogonal, called texts. Some basic theorems on enscribable texts are given, and a relationship to probabilistic quantum cloning is demonstrated.

\end{abstract}


\maketitle

We beg the reader's indulgence as we quickly move through some basic definitions. A quantum {\it N-text} ${T=\{\ket{\psi_1},\dots ,\ket{\psi_N}\}}$ is a set of $N$ (normalized) states $\ket{\psi_i}$ in the Hilbert space $\cal H$ of a quantum system, with no two being colinear. Such a text~$T$ is called {\it classical} if the states it contains are mutually orthogonal, {\it fully-quantum} if {\it no} two states in $T$ are orthogonal, and {\it efficient} if the states in $T$ are linearly independent.~(Note that all classical texts and all 2-texts are efficient.) $\cal H$ is called the {\it language} of $T$, and the subspace ${{\cal H}_T={\rm span}\,(\ket{\psi_1},\dots ,\ket{\psi_N})\subseteq{\cal H}}$ the {\it dialect} of $T$. If ${\cal H}_T={\cal H}$, then $T$ is said to be {\it thick} --- otherwise it is {\it thin}. (There is also the obvious notion of a {\it subtext} of~$T$.) Two $N$-texts $T=\{\ket{\psi_i}\}$ and $T^{\prime}=\{\ket{\psi^{\prime}_i}\}$ with the same language $\cal H$ are {\it equivalent} if there exist complex numbers $\beta_i$ of modulus 1, a unitary operator $V$ on $\cal H$, and a permutation $\pi$ of $\{1,\dots ,N\}$ such that $\ket{\psi^{\prime}_i}=\beta_i\,V\ket{\psi_{\pi(i)}}$ for all $1\leq i\leq N$; this defines an equivalence relation on all $N$-texts with language $\cal H$. Whether or not a text is classical, fully-quantum, efficient, or thick, is invariant under equivalence.

Now consider a composite system with Hilbert space ${\cal H}\otimes{\cal H}$, along with the $N$ entangled bipartite states $\{\ket{\Omega^T_i(q,\psi_0)}\}$ constructed from the $N$-text ${T=\{\ket{\psi_i}\}}$ (with language $\cal H$) as 
\begin{equation}
\ket{\Omega^T_i(q,\psi_0)}={1\over\sqrt{A_i}}\big(\ket{\psi_i}\otimes\ket{\psi_0}+q\ket{\psi_0}\otimes\ket{\psi_i}\big), \label{on}
\end{equation}
for some fixed unit vector $\ket{\psi_0}$ in $\cal H$ and complex number~$q$, where $A_i=1+|q|^2+2\,{\rm Re}(q)\,|\langle\psi_i|\psi_0\rangle|^2$. We will say that $T$ can be {\it q-enscribed} if there exists a state $\ket{\psi_0}$ in $\cal H$ and a unitary transformation $U$ on ${\cal H}\otimes{\cal H}$ such that 
\begin{equation}
U\ket{\Omega^T_i(q,\psi_0)}=\alpha_i\ket{\psi_i}\otimes\ket{\psi_i} \label{th}
\end{equation}
for each $1\leq i\leq N$ and for some complex numbers $\alpha_i$ of modulus one. In this case, $\ket{\psi_0}$ is called a {\it q-tablet} for $T$, $U$ is called the {\it procedure} of the {\it q-enscription of $T$ onto $\ket{\psi_0}$}, and the $\alpha_i$'s are referred to as the {\it output phases} of the $q$-enscription. Moreover, we simply say that a text can be {\it enscribed} if it can be $q$-enscribed for some~$q$. Similarly, $\ket{\psi_0}$ is a {\it tablet} for $T$ if it is a $q$-tablet for $T$ for some $q$. (Note that a $q$-enscription of $T$ provides a $q$-enscription of every subtext of $T$). A $q$-enscription of an $N$-text ${T=\{\ket{\psi_i}\}}$ is called {\it weakly central} if the associated $q$-tablet satisfies $|\langle \psi_i|\psi_0\rangle|=|\langle \psi_j|\psi_0\rangle|$ for all ${1\leq i,j\leq N}$. In this case we call $\ket{\psi_0}$ a {\it weakly central q-tablet} for $T$. If the tablet obeys the stronger condition $\langle \psi_i|\psi_0\rangle=\langle \psi_j|\psi_0\rangle$ for all $1\leq i,j\leq N$, then the enscription (and tablet) are simply called {\it central}. Finally, an enscription and tablet are said to be {\it quasi-central} if exactly $N-1$ of the quantities $\langle \psi_i|\psi_0\rangle$ are equal.

The case $q=0$ in (\ref{th}) reduces to what is commonly called {\it deterministic cloning}. In our terminology, the well-known No-Cloning Theorem~\cite{foot1} simply states: {\it A text T can be 0-enscribed if and only if T is classical}. The purpose of this paper is to understand how this result changes as a function of $q$. But first we will need a more manageable characterization of enscription. 

The unitary transformation $U$ in (\ref{th})  exists if and only~if 
\begin{equation}
\langle \Omega^T_i(q,\psi_0)|\Omega^T_j(q,\psi_0)\rangle ={\overline \alpha_i}\alpha_j\langle \psi_i|\psi_j\rangle^2 \label{fo}
\end{equation}
for all $1\leq i<j\leq N$. (Note that (\ref{fo}) is automatically satisfied when $i=j$.) These ${N(N-1)\over 2}$ conditions can be rewritten as
\begin{equation}
z_{ij}+Q\,\langle \psi_i|\psi_0\rangle\langle\psi_0|\psi_j\rangle = \sqrt{B_iB_j}\,\gamma_{ij}\,z_{ij}^2\ \ \ , \label{fi}
\end{equation}
where $z_{ij}=\langle \psi_i|\psi_j\rangle$, $Q={2\,{\rm Re}(q)\over 1+|q|^2}$, $B_i=1+Q\,|\langle \psi_i|\psi_0\rangle|^2$, and $\gamma_{ij}={\overline \alpha_i}\alpha_j$. The $N\times N$ matrix defined by the quantities $z_{ij}$ is known as the Gram matrix of $T$. The real number $-1\leq Q\leq 1$ is called the {\it entanglement parameter} of the enscription, and satisfies $Q=1$ (respectively, $Q=-1$) if and only if $q=1$ (respectively, $q=-1$). We see from (\ref{fi}) that if the complex numbers $q_1$ and $q_2$ lead to the same value of $Q$, then a given text $T$ can be $q_1$-enscribed if and only if it can be $q_2$-enscribed. This immediately gives our first result, which is a simple generalization of the No-Cloning Theorem

\vskip 5pt
\noindent
{\bf Theorem~1:} If ${\rm Re}(q)=0$, then a text $T$ can be \mbox{$q$-enscribed} if and only if $T$ is classical. 

\vskip 5pt
\noindent
(Moreover, for $q$ and $T$ as in Theorem~1, it is easy to show that {\it every} $\ket{\psi_0}$ in $\cal H$ is a $q$-tablet for $T$.) Note that the property of a text which determines whether or not it can be $q$-enscribed with ${\rm Re}(q)=0$, namely classicality, is invariant under textual equivalence. More generally, for any $q$ we have that the $q$-enscribability of a text is invariant under equivalence:
\pagebreak[4]
\vskip 5pt
\noindent
{\bf Theorem~2:} Let $T=\{\ket{\psi_i}\}$ and $T^{\prime}=\{\ket{\psi^{\prime}_i}\}$ be equivalent $N$-texts with $\ket{\psi^{\prime}_i}=\beta_i\,V\ket{\psi_{\pi(i)}}$. Then $T$ can be $q$-enscribed (with procedure $U$, tablet $\ket{\psi_0}$ and output phases $\alpha_i$) if and only if $T^{\prime}$ can be $q$-enscribed (with procedure $U^{\prime}=(V\otimes V)\,U\,(V\otimes V)^{-1}$, tablet $\ket{\psi^{\prime}_0}=V\ket{\psi_0}$, and output phases $\alpha^{\prime}_i=\alpha_{\pi(i)}{\overline \beta_i}$). 

\vskip 3pt
\noindent
{\it Proof:} This follows immediately from (\ref{th}) and (\ref{fi}).~{\it Q.E.D.}

\vskip 5pt
\noindent
Note also that an enscription of a text $T$ is weakly central if and only if the associated enscription (as in Theorem~2) of an equivalent text $T^{\prime}$ is weakly central. (The same does not hold if we replace ``weakly central'' by ``central''.) Another consequence of Theorem~2 is that any $q$-enscribable text $T$ is equivalent to a text $T^{\prime}$ which can be $q$-enscribed with trivial output phases. However, there do exist texts for which every $q$-enscription requires non-trivial output phases. We will give examples later.

Besides Theorem~1, classical texts have various other nice properties. For instance
\vskip 5pt
\noindent
{\bf Theorem~3:} A text $T$ can be $q$-enscribed for {\it every} $q$ if and only if $T$ is classical.

\vskip 3pt
\noindent
{\it Proof:} The ``only if'' follows immediately from Theorem~1. For the ``if'', simply choose the tablet $\ket{\psi_0}$ to be any one of states in $T$. {\it Q.E.D.}

\vskip 5pt
\noindent
It is easy to see that if ${\rm Re}(q)\neq 0$, then $\ket{\psi_0}$ in $\cal H$ is a $q$-tablet for a classical $N$-text $T$ if and only if it is orthogonal to (at least) $N-1$ of the vectors in $T$ (such as for the tablet chosen in the above proof). Thus, in this case, not every $\ket{\psi_0}$ in $\cal H$ is a $q$-tablet for $T$. (This feature persists for a generic enscribable text.) In particular, for $Q\neq 0$, all enscriptions of a thick classical text are quasi-central, while the thin case also allows central enscriptions. We now show that for $Q\neq 0$ there are many enscribable texts which are not classical.

\vskip 5pt
\noindent
{\bf Theorem~4:} All 2-texts can be enscribed.
\vskip 3pt
\noindent
{\it Proof:} By Theorem~2 we may assume that ${0\leq z_{12}<1}$, since the equivalence class of a 2-text is completely determined by $|z_{12}|$. Now choose the central tablet ${\ket{\psi_0}=(\ket{\psi_1}+\ket{\psi_2})/\sqrt{2\,(1+z_{12})}}$, and let ${Q=-2z_{12}/(1+z_{12})^2}$ and $\alpha_i=1$ (for all $i$). It is easy to check that (\ref{fi}) is satisfied, and $|Q|<1$. {\it Q.E.D.}

\vskip 5pt
\noindent
A characterization of the values of $Q$ associated with enscriptions of any fixed 2-text will be given in Theorem~11. Our next result provides us with a large class of texts which cannot be $q$-enscribed for {\it any} $q$. (We will call such texts {\it illegible}.)

\vskip 5pt
\noindent
{\bf Theorem~5:} If a text $T$ can be enscribed, then $T$ is efficient.

\vskip 5pt
\noindent
{\it Proof:} We demonstrate the result for 3-texts. The generalization to $N>3$ is straightforward. Assume that the text $T=\{\ket{\psi_1},\ket{\psi_2},\ket{\psi_3}\}$ is enscribable with procedure $U$, and that ${\ket{\psi_3}=\lambda_1\ket{\psi_1}+\lambda_2\ket{\psi_2}}$. We know that $U\ket{\Omega^T_3}$ is proportional both to $\ket{\psi_3}\otimes\ket{\psi_3}$ (by (\ref{th})) and to a linear combination of $\ket{\psi_1}\otimes\ket{\psi_1}$ and $\ket{\psi_2}\otimes\ket{\psi_2}$ (by the linearity of $U$ and (\ref{th})). But this requires that either $\lambda_1$ or $\lambda_2$ is zero, contradicting the assumption that $T$ is a text (since $\ket{\psi_1}$ is then colinear with $\ket{\psi_2}$ or $\ket{\psi_3}$).~{\it Q.E.D.}

\vskip 5pt
\noindent
In particular, there exist 3-texts which are not enscribable. Theorem~5 also implies that an enscribable text can be probabilistically cloned, since it is known~\cite{foot2} that (in our terminology) a text can be probabilistically cloned if and only if it is efficient. A natural question then is whether the converse of Theorem~5 is true. Our next theorem shows that the answer is ``no", but first we will need the following two lemmas and a definition.

\vskip 5pt
\noindent
{\bf Lemma~1:} Let $\ket{v_1}$, $\ket{v_2}$ and $\ket{v_3}$ be unit vectors in a complex inner-product space. Then $|\langle v_1|\,v_2\rangle|^2 + |\langle v_2|\,v_3\rangle|^2 + |\langle v_3|\,v_1\rangle|^2\leq 1 + 2\, {\rm Re}[\langle v_1|\,v_2\rangle\langle v_2|\,v_3\rangle\langle v_3|\,v_1\rangle]$. 

\vskip 3pt
\noindent
{\it Proof:} The inequality is violated if and only if the Gram matrix of the set $\{\ket{v_1},\ket{v_2},\ket{v_3}\}$, which must be positive semi-definite~\cite{foot2}, has negative determinant. {\it Q.E.D.}

\vskip 5pt
\noindent
{\bf Lemma~2:} Let $\ket{\psi_0}$ be a $q$-tablet for an enscribable text~$T$, with ${\rm Re}(q)\neq 0$. Then for any fixed $i$~and~$j$ (${i\neq j}$), we have $z_{ij}=0$ if and only if ${\langle\psi_0|\psi_i\rangle\langle\psi_0|\psi_j\rangle =0}$. 

\vskip 5pt
\noindent
{\it Proof:} The ``only if'' implication follows trivially from (\ref{fi}), as does the ``if'' implication if {\it both} $\langle\psi_0|\psi_i\rangle =0$ {\it and} ${\langle\psi_0|\psi_j\rangle =0}$. If only one of these is true, say $\langle\psi_0|\psi_j\rangle =0$, then we obtain from (\ref{fi}) that either ${z_{ij}=0}$ or $|z_{ij}|^2={1\over B_i}$. In the latter case we must have $Q>0$ since $|z_{ij}|<1$. Also in this case the inequality in Lemma~1 for the vectors $\{\ket{\psi_0},\ket{\psi_i},\ket{\psi_j}\}$ reads ${|\langle\psi_0|\psi_i\rangle|^2+{1\over 1+Q\,|\langle\psi_0|\psi_i\rangle|^2}\leq 1}$, and it is straightforward to show that this imples ${|\langle\psi_0|\psi_i\rangle|^2<0}$. {\it Q.E.D.}

\vskip 5pt
\noindent
We will say that a text $T$ is a {\it direct sum} of the subtexts $T_1=\{\ket{\psi_1},\dots ,\ket{\psi_{N_1}}\}$ and $T_2=\{\ket{\phi_1},\dots ,\ket{\phi_{N_2}}\}$ if ${T=\{\ket{\psi_1},\dots ,\ket{\psi_{N_1}},\ket{\phi_1},\dots ,\ket{\phi_{N_2}}\}}$ and $\langle\psi_i|\phi_j\rangle =0$ for all $1\leq i\leq N_1$ and $1\leq j\leq N_2$. (There is an obvious generalization to a direct sum of finitely many texts.) We can now prove a sharper version of Theorem~5.

\vskip 5pt
\noindent
{\bf Theorem~6:} Every enscribable text $T$ is a direct sum~of a classical subtext and an efficient fully-quantum subtext.

\vskip 3pt
\noindent
{\it Proof:} Let $\ket{\psi_0}$ be a tablet for $T$. Define the subtext $T_1$ of $T$ to be the set of all states in $T$ which are orthogonal to $\ket{\psi_0}$, and the subtext $T_2$ to contain the remaining states in $T$. It follows from Lemma~2 that $T_1$ is classical, $T_2$~is fully-quantum, and $T$ is a direct sum of $T_1$ and $T_2$. Finally, Theorem~5 shows that $T_2$ is efficient. {\it Q.E.D.}

\vskip 5pt
\noindent
In particular, there exists a large class of efficient \mbox{3-texts} that cannot be enscribed --- namely, those with exactly one Gram matrix element $z_{ij}$ ($i<j$) which is zero. (There are, of course, similar examples for $N>3$.) Such illegible, efficient texts can still be probabilistically  cloned. Thus, in this sense, enscription is ``in-between" deterministic and probabilistic cloning. We will say more on this relationship shortly.

We can now ask whether the converse of Theorem~6 is true, which would give us a complete classification of enscribable texts. Some hope may be garnered from

\vskip 5pt
\noindent
{\bf Theorem~7:} Let $T$ be a direct sum of a classical subtext $T_1$ and an enscribable subtext $T_2$. Then $T$ is enscribable.

\vskip 3pt
\noindent
{\it Proof:} Note that ${\cal H}_T={\cal H}_{T_1}\oplus{\cal H}_{T_2}$. Without loss of generality we assume $T$ is thick, so that ${\cal H}_T={\cal H}$. Let $P$ be the projection operator onto ${\cal H}_{T_2}$. If $T_2$ is enscribable with procedure~$U$, tablet $\ket{\psi_0}$, and entanglement parameter~$Q$, then there exists a unitary operator $V$ on ${\cal H}_{T_1}$ such that $T$ can be enscribed with procedure $U^{\prime}={V\oplus (P\,UP)}$, tablet $\ket{\psi^{\prime}_0}=P\ket{\psi_0}/||P\ket{\psi_0}||$, and $Q^{\prime}=||P\ket{\psi_0}||^2Q$. The existence of $V$ is guaranteed by Theorem~3 and the comment that follows it. {\it Q.E.D.}

\vskip 5pt
\noindent
Thus, the status of the converse of Theorem~6 is reduced to the question of whether or not all efficient, fully-quantum texts can be enscribed. We now study a group of texts which shows that the answer is ``no'', dashing our hopes of a simple classification of enscribable texts.

A text $T$ is called {\it real} if all of its Gram matrix elements~$z_{ij}$ are real, and {\it uniform} if there exists~a~\mbox{complex} number $z$ such that $z=z_{ij}$ for all $i\neq j$. The value of $z$ for a uniform {\it and} real text $T$ must satisfy $-{1\over N-1}\leq z<1$ (with $z\neq -{1\over N-1}$ if $T$ is efficient.) The following result, which will be proven in~\cite{foot3}, gives (along with Theorem~4) a classification of all real, uniform texts that can be enscribed. It can also be seen as a corollary of Theorem~12 below, which characterizes the values of~$Q$ associated with enscriptions of any given real, uniform $N$-text with $N\geq 3$.

\vskip 5pt
\noindent
{\bf Theorem~8:} A real, uniform $N$-text $T=\{\ket{\psi_i}\}$, $N\geq 3$, can be enscribed if and only if $z\geq z_0$, where $z=\langle\psi_i|\psi_j\rangle$ ($i\neq j$) and $-{1\over N-1}<z_0<0$ satisfies $f(z_0)=0$, with $f(z)=1+2\,(N-1)z-{3\,(N-2)z^2}+4\,(N-1)z^3+3\,(N-2)z^4-2\,(2N-3)z^5+(N-1)z^6$. (For any fixed~$N$, $z_0$ is the unique solution of $f(z)=0$ in this range.)

\vskip 5pt
\noindent
As an example, for $N=3$ we have $z_0\simeq -.203785$, so that real, uniform 3-texts with $-0.5<z<z_0$ are illegible. (At large $N$, we have $z_0\simeq -{1\over 2N}$.) Since real, uniform $N$-texts (with $z\neq 0,-{1\over N-1}$) are efficient and fully-quantum, we see that the converse of Theorem~6 is false. 

Additional classes of illegible texts are provided by the next theorem. Its proof, which requires techniques substantially different from those developed here, will also be given in~\cite{foot3}.

\vskip 5pt
\noindent
{\bf Theorem~9:} Let $T$ be an enscribable, fully-quantum $N$-text ($N\geq 3$) with Gram matrix $z_{ij}$. Then the matrix  $M_{ij}={1\over z_{ij}}$ has a nonzero determinant, and (exactly) $N-1$ of the eigenvalues of $M$ have the same sign ``$\epsilon\,$''. Moreover, ${{\rm sign}(Q)=\epsilon}$ for any enscription of $T$.

\vskip 5pt
\noindent
In particular, only for $N=2$ can a fully-quantum \mbox{$N$-text} possibly have enscriptions with both positive and negative $Q$ (and they always do, as we will see in Theorem~11).

Before describing the enscriptions of real, uniform texts in more detail, we turn to the relationship between enscriptions of thick and thin texts (refining the discussion in the proof of Theorem~7). Consider a thin text $T$ with language $\cal H$. One can always write ${\cal H}={\cal H}_T\oplus{\cal H}_{\perp}$, where ${\cal H}_{\perp}$ is the subspace of~$\cal H$ orthogonal to the dialect ${\cal H}_T$. Assume that we have an enscription of $T$ --- with procedure $U$, output phases $\alpha_i$, and entanglement parameter $Q_0$ --- such that the tablet $\ket{\psi_0}$ lies in ${\cal H}_T$. In this case we can assume, without loss of generality, that the procedure has the form $U=V\oplus I$, where $V$ is a unitary operator on ${\cal H}_T$ and $I$ is the identity operator on~${\cal H}_{\perp}$. Thus, we can view the above as an enscription of~$T$ considered as a {\it thick} text with language ${\cal H}_T$, and having procedure $V$. We can now use this to generate a family of enscriptions of the thin text~$T$ with any of the entanglement parameters $Q(t)={1\over t}\,Q_0$, where $|Q_0|\leq t\leq 1$. Simply keep the same procedure $U$ and output phases $\alpha_i$ as above, but choose the one-parameter family of tablets $\ket{\psi_0(t)}=\sqrt{t}\ket{\psi_0}+\sqrt{1-t}\ket{\phi_0}$, where the unit vector~$\ket{\phi_0}$ lies in ${\cal H}_{\perp}$. The restriction on the range of $t$ simply insures that $|Q(t)|\leq 1$. Thus, if we are interested in the range of $Q$'s associated with enscriptions of a given text~$T$, it suffices to consider the thick case. The lower bound for $|Q|$ is the same in both the thick and thin cases, while the upper bound in the thin case is always $|Q|=1$. Indeed, for a thick text one can never reach $Q=-1$.

\vskip 5pt
\noindent
{\bf Theorem~10:} A thick $N$-text $T$ cannot be enscribed with $Q=-1$.

\vskip 3pt
\noindent
{\it Proof:} When $Q=-1$ and $T$ is thick, we have that the $\ket{\Omega_i(-1,\psi_0)}$'s are linearly dependent. But if $T$ is enscribable, we have (by Theorem~5) that the $\ket{\psi_i}$'s, and hence the ${\ket{\psi_i}\otimes\ket{\psi_i}}$'s, are linearly independent. {\it Q.E.D.}

\vskip 5pt
\noindent
We now state results, which will be proven in~\cite{foot3}, that give the range of entanglement parameters for enscriptions of 2-texts and real, uniform $N$-texts ($N\geq 3$).

\vskip 5pt
\noindent
{\bf Theorem~11:} A thick 2-text $T=\{\ket{\psi_1},\ket{\psi_2}\}$, with ${z=\langle\psi_1|\psi_2\rangle}$, can be enscribed for ${{2\,|z|\over 1+|z|^2}\leq Q\leq 1}$ and $-1< Q\leq -{2\,|z|\over (1+|z|)^2}$, and for no other values of $Q$. The boundaries $Q={2\,|z|\over 1+|z|^2}$ and $Q=-{2\,|z|\over (1+|z|)^2}$ are associated with weakly central enscriptions.

\vskip 5pt
\noindent
As an example, we give a concrete realization of a central 1-enscription procedure $U$ for the (thick) $2$-text $\{\ket{\psi_1}, \ket{\psi_2}\}$, with $z=\langle\psi_1|\psi_2\rangle=\sqrt{3}-2$. One can view this as a qubit system, where in the (orthonormal) ``computational basis'' $\{\ket{0},\ket{1}\}$ we have $\ket{\psi_1}= a_+\ket{0}+a_-\ket{1}$ and $\ket{\psi_2}=a_+\ket{0}-a_-\ket{1}$, with $a_{\pm}=\sqrt{(1\pm z)/2}$. If we choose the central tablet $\ket{\psi_0}=\ket{0}$, then the unitary operator which accomplishes the 1-enscription (with trivial output phases) is $U = \frac{1}{2}(\ket{00}\bra{00}-\ket{11}\bra{11})+{\frac{\sqrt{3}}{2}(\ket{00}\bra{11}+\ket{11}\bra{00})}+(\ket{10}\bra{10}+\ket{01}\bra{01})$, in the standard notation ${\ket{ab}=\ket{a}\otimes\ket{b}}$. 

One nice feature of this example is that it can be interpreted as the evolution of a pair of {\it indistinguishable} systems obeying {\it Bose statistics}, since not only are the initial and final states symmetric under the ``permutation operator'' $P$ exchanging the two copies of $\cal H$ in the Hilbert space ${\cal H}\otimes{\cal H}$, but we also have that the enscription procedure $U$ commutes with $P$. (This condition can always be imposed for a 1-enscribable text.) Thus, for example, the computational basis states may be viewed as two orthogonal polarization states of a single photon, and $U$ the evolution operator of a two-photon system. For such bosonic systems, the initial entanglement needed for \mbox{1-enscription} comes for ``free''. For other systems it has to be generated by some process, such as the one in our discussion of probabilistic cloning below. But first, we give a generalization of Theorem~11 to $N\geq 3$.

\vskip 2pt
Defining the quantities $Q_2=-\frac{Nz}{(1+z)\,[1+(N-1)z]}$~and ${Q_1=-\frac{z\,[4z (1-z^2) (1-z) + N (1 - 2z + 3z^2 + 2z^3 - 3z^4)]}{(1+z)\,[1+(N-1)z]\,(1-z+z^2)^2}}$,~we~have

\vskip 7pt
\noindent
{\bf Theorem~12:} For $N\geq 3$, a thick real, uniform \mbox{$N$-text} $T=\{\ket{\psi_i}\}$, with ${z=\langle\psi_i|\psi_j\rangle}$ ($i\neq j$), can be enscribed for every $Q$ in the interval $[Q_1,Q_2]\,\cap\,(-1,1]$, and for no other values of~$Q$. Moreover, the value $Q=Q_2$ is associated with the unique (up to an overall phase) central tablet for $T$, and all other enscriptions of $T$ are quasi-central.

\vskip 5pt
\noindent
Note that we always have ${\rm sign}(Q)=-{\rm sign}(z)$, as is required by Theorem~9, and that the condition $Q_1\leq 1$ leads to Theorem~8. One can also show that the enscriptions in Theorem~12 with $Q\neq Q_1,Q_2$ {\it require} non-trivial output phases.

Finally, we show how enscription can be used as a resource for probabilistic cloning. Consider a $q$-enscribable text $T$, along with a $q$-tablet $\ket{\psi_0}$. If there is a way to obtain the set $\{\ket{\Omega^T_i(q,\psi_0)}\}$ {\it probabilistically} from the set $\{\ket{\psi_i}\otimes\ket{\psi_0}\}$ (using a combination of unitary evolution and measurement), then combining this process with the appropriate $q$-enscription procedure $U$ results in a machine that probabilistically clones $T$. For example, the following such ``quantum operation'' does the job. First, introduce an auxiliary qubit system in the pure state ${\ket{\xi_q} = \frac{1}{\sqrt{2}}\Big(\ket{0} + \frac{q}{|q|}\ket{1}\Big)}$. Next, operate on ${\ket{\xi_q}\otimes\ket{\psi_i}\otimes\ket{\psi_0}}$ with the unitary ``controlled-swap'' $S$  that acts as (see \cite{foot1})
\begin{eqnarray*}
S\,(\ket{0}\otimes\ket{\psi}\otimes\ket{\phi}) =& \ket{0}\otimes\ket{\psi}\otimes\ket{\phi}\nonumber\\
S\,(\ket{1}\otimes\ket{\psi}\otimes\ket{\phi}) =& \ket{1}\otimes\ket{\phi}\otimes\ket{\psi}
\end{eqnarray*}
for all $\ket{\psi}$ and $\ket{\phi}$ in $\cal H$. This yields 
\begin{displaymath}
\sqrt{p_i}\ket{\eta_q}\otimes\ket{\Omega_i(q,\psi_0)} +   \sqrt{1 - p_i}\,\ket{\chi_q}\otimes\ket{\Omega_i(-q/|q|,\psi_0)},\label{st}
\end{displaymath}
where the states $\ket{\eta_q}=\frac{1}{\sqrt{1 + |q|}}\Big( \ket{0}+\sqrt{|q|}\,\ket{1}\Big)$ and $\ket{\chi_q}=\frac{1}{\sqrt{1 + |q|}} \Big(\sqrt{|q|}\,\ket{0}-\ket{1}\Big)$ are orthogonal, and $p_i=\frac{A_i}{(1+|q|)^2}$. If we now measure the observable ${(\ket{\eta_q}\bra{\eta_q})\otimes I\otimes I}$ --- with eigenvalues $\lambda =1,0$ --- we will know whether our process has been ``successful'' or not. The probability of success ($\lambda =1$) is equal to $p_i$, which simplifies in the case of real $q$ to
\begin{displaymath}
p_i=\frac{1+Q\,|\langle\psi_i|\psi_0\rangle|^2}{1+|Q|}.
\end{displaymath}
If our measurement indicates success, we then operate on the resulting state $\ket{\eta_q}\otimes\ket{\Omega_i(q,\psi_0)}$ with~${I\otimes U}$ to obtain ${\alpha_i\ket{\eta_q}\otimes\ket{\psi_i}\otimes\ket{\psi_i}}$ (see (\ref{th})). We can now discard the auxiliary system.

While there are many quantum operations that produce the $\ket{\Omega^T_i (q, \psi_0)}$'s with varying probabilities, the one chosen here is particularly simple. Also, for $N=2$ and $-1<z=z_{12}<0$ (which is always true up to equivalence), the above probabilities saturate the upper bound $p_i=1/(1+|z|)$ for probabilistic cloning given in \cite{foot2} if we use the central enscription of $T$ (which has $Q=-2\,z/(1 + z^2)$, $|\langle\psi_i|\psi_0\rangle|^2=(1+z)/2$, and output phases satisfying ${\overline \alpha_1}\alpha_2=-1$). Another interesting feature of the above probabilistic cloning machine is that the ``failure states" $\ket{\Omega_i(-q/|q|,\psi_0)}$ may be useful for other purposes since they are (for real $q$) either symmetrizations (if $Q<0$) or anti-symmetrizations (if~$Q>0$) of each of the states in $T$ with the state $\ket{\psi_0}$. We develop these and related ideas further~in~\cite{foot3}.

What remains to be done? Of course, the biggest open question is: {\it Which efficient, fully-quantum texts can be enscribed}? More ambitiously, one could hope for a parameterization of all enscriptions of any given text $T$ (that is, all associations of $q$, tablet, procedure, and output phases). We will continue our discussion of these issues in~\cite{foot3}. However there is a simple generalization of enscription, called {\it translation}, for which we {\it have} been able to find such a complete classification --- and the statement and proof of this result displays deep connections to several areas in pure mathematics. This will be the topic of the companion paper to this letter~\cite{foot4}.

We thank Klaus Bering and Mark Mueller for important contributions.
This work was supported in part by the U.S. Department of Energy.


\begin{thebibliography}{99}
\bibitem{foot1} M.~A.~Nielsen and I.~L.~Chuang, {\it Quantum Computation and Quantum Information} (Cambridge University Press, Cambridge, 2000), and references therein.
\bibitem{foot2} L.-M.~Duan and G.-C.~Guo, Phys. Rev. Lett. {\bf 80} (1998) 4999.
\bibitem{foot3} K.~Bering, R.~Espinoza, T.~Imbo, P.~Lopata and M.~Mueller, in preparation.
\bibitem{foot4} R.~Espinoza, T.~Imbo and P.~Lopata, quant-ph/0403210.

\end{thebibliography}
\end{document}